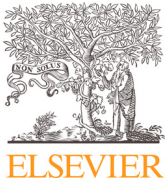
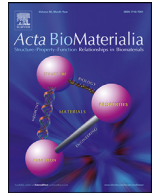

# Evaluation of the penetration process of fluorescent collagenase nanocapsules in a 3D collagen gel

Víctor M. Moreno[a,b], Alejandro Baeza[c,*], María Vallet-Regí[a,b,*]

[a] *Dpto. Química en Ciencias Farmacéuticas, Universidad Complutense de Madrid, Instituto de Investigación Sanitaria, Hospital 12 de Octubre i+12, Plaza Ramón y Cajal s/n, Spain*
[b] *CIBER de Bioingeniería, Biomateriales y Nanomedicina, CIBER-BBN, 28040 Madrid, Spain*
[c] *Dpto. Materiales y Producción Aeroespacial, ETSI Aeronáutica y del Espacio, Universidad Politécnica de Madrid, 28040 Madrid, Spain*

## ABSTRACT

One of the major limitations of nanomedicine is the scarce penetration of nanoparticles in tumoral tissues. These constrains have been tried to be solved by different strategies, such as the employ of polyethyleneglycol (PEG) to avoid the opsonization or reducing the extracellular matrix (ECM) density. Our research group has developed some strategies to overcome these limitations such as the employ of pH-sensitive collagenase nanocapsules for the digestion of the collagen-rich extracellular matrix present in most of tumoral tissues.

However, a deeper understanding of physicochemical kinetics involved in the nanocapsules degradation process is needed to understand the nanocapsule framework degradation process produced during the penetration in the tissue. For this, in this work it has been employed a double-fluorescent labelling strategy of the polymeric enzyme nanocapsule as a crucial chemical tool which allowed the analysis of nanocapsules and free collagenase during the diffusion process throughout a tumour-like collagen matrix. This extrinsic label strategy provides far greater advantages for observing biological processes. For the detection of enzyme, collagenase has been labelled with fluorescein Isothiocyanate (FITC), whereas the nanocapsule surface was labelled with rhodamine Isothiocyanate (RITC). Thus, it has been possible to monitor the hydrolysis of nanocapsules and their diffusion throughout a thick 3D Collagen gel during the time, obtaining a detailed temporal evaluation of the pH-sensitive collagenase nanocapsule behaviour. These collagenase nanocapsules displayed a high enzymatic activity in low concentrations at acidic pH, and their efficiency to penetrate into tissue models pave the way to a wide range of possible nanomedical applications, especially in cancer therapy.

**Statement of significance**

The present study is focused on the development of a dual fluorescent labelling strategy that allow to monitor the penetration of pH-cleavable polymeric nanocapsules through three-dimensional tissue models. Collagenase was housed within the nanocapsules which were engineered to be disassembled at acidic pH leading to enzyme release. In this work, the pH-responsive behaviour of these nanocapsules has been studied employing 3D collagen matrices as tissue models, and the results have confirmed a significant penetration and more homogeneous distribution of the proteolytic enzymes in mild-acidic conditions. This work presents important applications in the field of cancer therapy and the treatment of fibrotic diseases, among others.

© 2020 Published by Elsevier Ltd on behalf of Acta Materialia Inc.

## 1. Introduction

Nanomedicine has reached great interest in the current decade through the development of nanoparticles and biological nanostructures for molecular diagnostics, treatment of cancer and other diseases, and for both applications, as theranostic agents [1]. In the case of cancer therapy, one of the major limitations of the application of nanomedicines is their scarce penetration in tumours [2,3]. This weakness is caused as consequence of the presence of several well-known biological barriers, that limits the efficacy of the nanomedicine drugs [4]. One of these limitation is the off-target accumulation of nanomedicines along the body which can be overcome by anchoring targeting moieties such as peptides or antibodies, among others, on nanoparticle surface [5]. Another constrain is the opsonization of the nanotherapeutic by the Mononuclear Phagocyte System (MPS) once it is injected in the bloodstream. Plasma proteins are adsorbed onto nanoparticles, forming a protein corona on the surface that undergoes their recognition by unspecific receptors of phagocytes, which engulf and destroy the nanotherapeutics. It has been widely reported that the surface decoration with polyethyleneglycol (PEG) [6] reduces this problem hiding the surface to the opsonins present in the blood stream. However, once the nanoparticle reaches the tissue, the existence in most of tumoral tissues of poor lymphatic drainage and dense extracellular matrix (ECM) generates high interstitial fluid pressure that provokes the extravasation of nanotherapeutics to distal regions. Different types of nanosystems which present characteristics that partially solve some of these limitations have been developed employing an extensive list of materials of different nature: from inorganic systems as metallic [7] and ceramic particles [8,9], organic ones like polymersomes, [10] micelles, [11] liposomes [12] and polymer nanocapsules [13] to hybrid nanodevices which combines both natures as protein nanocapsules [14] and even bio-hybrid nanocarriers with living organisms as bacteria [15]. The myriad of nanosystems developed is endless, but unfortunately, many of them fail in their goal [16].

* Corresponding authors.
*E-mail addresses:* alejandro.baeza@upm.es (A. Baeza), vallet@ucm.es (M. Vallet-Regí).

Our research group has reported some strategies to improve the penetration of nanomedicines in tumoral tissues. One of these strategies is the development of pH-sensitive collagenase nanocapsules [17] for the digestion of the collagen-rich ECM present in tumoral tissues, yielding an enhanced penetration of nanoparticles in tumoral tissue models. These collagenase nanocapsules have been used for many clinical applications both in cancer [18] and fibrosis [19]. The knowledge of physicochemical kinetic involved in the nanocapsules degradation process can help us to better understand the nanocapsule framework degradation process produced during the penetration in the tissue. This knowledge would allow to improve the nanocapsule performance for different therapeutic applications. Thus, in this work we describe the development of a fluorescent labelling strategy of the collagenase nanocapsules as a crucial chemical tool that allows the analysis of the diffusion mechanisms across the tissue [20]. This extrinsic labels provide far greater advantages for observing biological processes. This strategy has allowed us to achieve a detailed temporal analysis and evaluation of the collagenase nanocapsule cleavage process to better understand the free collagenase enzyme diffusion throughout a tumour-like collagen matrix. For the detection of enzyme, we have labelled Collagenase with Fluorescein Isothiocyanate (FITC), whereas the nanocapsule surface was labelled with Rhodamine Isothiocyanate (RITC). By this way, we could follow during the time the hydrolysis of nanocapsules and the diffusion of collagenase throughout a thick 3D Collagen gel. This collagen matrix has been widely employed in the literature [21,22] as model of extracellular matrix which allows to study the behaviour of nanomedicines in biological tissues. The detailed analysis of the data obtained could clarify some of the factors involved in the process, such as the penetration capacity of collagenase in low concentrations, the influence of the pH on this capacity, and the temporal evolution in the penetration reached by collagenase and nanocapsules.

## 2. Experimental section

### 2.1. Materials

The chemicals were bought to the corresponding supplier and they have been used without further purification. Collagenase Type I from Life Technologies; Acrylamide (Aa) from Fluka; Sodium Bicarbonate ($NaHCO_3$), Sodium Phosphate ($Na_3PO_4$), 2-Aminoethyl methacrylate hydrochloride (Am), Ethylene glycol dimethacrylate (EG), Ammonium persulfate (APS), N,N,N',N'-Tetramethylethylenediamine (TMEDA), Fluorescein isothiocyanate isomer I (FITC), Rhodamine B isothiocyanate mixed isomers (RITC), Dimethylsulfoxide (DMSO), Paraformaldehyde and Sodium Hydroxide (NaOH) from Sigma Aldrich; Amicon® Ultra-2mL Centrifugal Filters Ultracel®- 10K from Millipore; Absolute Ethanol from Panreac; Dulbecco's modified Eagle's medium (DMEM), Fetal Bovine Serum (FBS), L-Glutamine and Antibiotic-Antimycotic (Anti-Anti) 100X from GIBCO. Rat tail Collagen (type I) and EnzChek$^{TM}$ Gelatinase/Collagenase Assay Kit from Life Technologies; 10X PBS Buffer solution pH 7.4 from Ambion.

Instrumental section: The hydrodynamic size of protein capsules was measured by means of a Zetasizer Nano ZS (Malvern Instruments) equipped with a 633 nm "red" laser. Transmission Electron Microscopy (TEM) was carried out with a JEOL JEM 3000 instruments operated at 300kV, equipped with a CCD camera. Sample preparation was performed by dispersing in distilled water and subsequent deposition onto carbon-coated copper grids. A solution of 1 % of phosphotungstic acid (PTA) pH 7.0 was employed as staining agent in order to visualize the protein capsules. Fluorescence was measure with Synergy 4, power supply for Biotek Laboratory Instrument 100-240VAC, 50/60Hz, 250W. Confocal microscope Leica SP-2 AOBS with digital camera Leica DFC 350 FX.

### 2.2. Fluorescein-labelling of collagenase (Col-FITC)

30 μL from a solution of FITC in DMSO (1 mg FITC per 100 μl DMSO) were added to a solution of Collagenase ($3.1 \times 10^{-5}$ mmol) in 1 mL of $NaHCO_3$ buffer (0.01 M, pH 8.5) under stirring. It was stirred during 2 hours at room temperature (RT) and protected from light. After that, the Collagenase solution was purified by centrifugal separation (3.000 rpm, 15 min) with a 10 kDa cut-of filter (AMICON Ultra-2 mL 10KDa) and washed five times with $NaHCO_3$ buffer (0.01 M pH 8.5). The enzyme was collected by centrifugation at 1.500 rpm for 5 min and diluted to a volume of 1 mL with $NaHCO_3$ buffer (0.01 M pH 8.5). It was stored at 4 °C protected from light.

### 2.3. Synthesis of fluorescein-collagenase nanocapsules (nCol-F)

Previous purified Fluorescein-Collagenase solution (Col-FITC) was deoxygenated by bubbling $N_2$ during at least 20 min. Meanwhile, $NaHCO_3$ buffer (0.01 M pH 8.5) was deoxygenated three times with freeze-vacuum-$N_2$ cycles, at room temperature. Then, 0.035 mmol of acrylamide (Aa), 0.026 mmol of 2-aminoethylmetacrylate hydrochloride (Am), and 0.01 mmol of ethyleneglycol dimetacrylate (EG) were dissolved in 1 mL of deoxygenated $NaHCO_3$ buffer (0.01 M pH 8.5) and were added to the solution of Col-FITC. This mixture was stirred at 300 rpm for 10 min under nitrogen atmosphere at room temperature. Then, 0.013 mmol of ammonium persulfate (APS) and 0.02 mmol of N, N, N′, N′-tetramethyl ethylenediamine (TMEDA) dissolved in 1 mL of the deoxygenated $NaHCO_3$ buffer (0.01 M pH 8.5) were added. The reaction mixture was stirred at 300 rpm for 90 min at room temperature under inert atmosphere. After this time, the encapsulated enzyme was purified by centrifugal separation with 10 kDa cut-of filters (AMICON Ultra-2 mL 10 KDa) and washed three times with $NaHCO_3$ buffer (0.01 M pH 8.5). These capsules of collagenase were diluted to a 1 mL volume and were preserved at 4 °C.

### 2.4. Rhodamine-labelling of collagenase nanocapsules (nCol-F-Rh)

To prepare red fluorescent Collagenase Nanocapsules, to the previous nanocapsules solution (nCol-F) was added 30 μL from a solution of RITC in DMSO (1 mg RITC per 100 μL DMSO). It was stirred during 2 hours at RT and protected from light. After that, the nanocapsules solution was purified by centrifugal separation (3.000 rpm, 15 min) with a 10 kDa cut-of filter (AMICON Ultra-2 mL 10KDa) and washed five times with $NaHCO_3$ buffer (0.01 M pH 8.5). The enzyme was collected by centrifugation at 1.500 rpm for 5 min and diluted to a volume of 1 mL with $NaHCO_3$ buffer (0.01 M pH 8.5). It was stored at 4 °C. Protein concentration was measured following the Bicinchoninic Assay (BCA) protocol for Protein quantification [23]. For this, 50 μL of the protein sample were added to 200 μL of a working solution containing Bicinchoninic acid and Cupric Sulphate. Mixture was incubated for 30 min at 40 °C and then absorbance was measured at 562 nm. The green and red fluorescence of *nCol-F-Rh* was determined at $\lambda_{abs}/\lambda_{em}$ of 495 nm/520 nm for FITC and $\lambda_{abs}/\lambda_{em}$ of 544 nm/576 nm for RITC, respectively.

### 2.5. Enzymatic activity measurements

The enzymatic activity of native collagenase (Col), Col-FITC, nCol-F and nCol-F-Rh was evaluated using and following the protocol EnzChek Gelatinase/Collagenase Assay Kit. For this experiment, 80 μL of Phosphate buffer 1X, 20 μL of Collagen-FITC, and 100 μL of each collagenase sample were used. All samples were standardized to an enzymatic activity of 0.2 U/mL in 1X PBS solution at pH 7 at room temperature and their enzymatic activity was studied at different times every 20 min, measuring the accumulated green fluorescence intensity in a fluorescence microplate reader ($\lambda_{abs} = 495$ nm / $\lambda_{em} = 515$ nm).

Enzymatic activity of Col and nCol-F-Rh samples tested at pH 5.5 and 37 °C was measured as described, but samples were in this case standardized to an enzymatic activity of 1.3 U/mL in 1X PBS solution.

### 2.6. Preparation of 3D collagen gels

Briefly, 2 ml of Rat Tail Collagen type I (3 mg•ml$^{-1}$) and 0.6 mL of complemented DMEM (DMEM medium with 10 % FBS and *L*-Glutamine) were mixed at 0 °C and subsequently 100 μL of a 2 M NaOH solution was added until neutral pH. Then, 0.5 ml of FBS and 1 ml of 1X PBS were added to the solution at 0 °C. Afterwards, 0.5 mL of this mixture was added to the wells of a 24-well plate and incubated at 37 °C at an atmosphere of 5 % $CO_2$ for 1 h to promote the collagen gelification. To avoid gel cracking, the edge of each well was encircled with a needle. Finally, 250 μL of PBS was added, and incubated at 37 °C at an atmosphere of 5 % $CO_2$ overnight. The resulting collagen gels were employed for the further experiments one day after gels formation.

### 2.7. Nanocapsules penetration evaluation in 3D collagen gels at 18 h end-point experiment

To the previously prepared 3D collagen gels, the supernatant was removed and 50 μL of a 0.089 mg/mL nanocapsules solution (theoretical activity of 2,57 U/mL) in 0.05 M $Na_3PO_4$ pH 7.2 buffer and 0.05 M $Na_3PO_4$ pH 5.5 buffer, respectively, was added in the centre of the gel. Next day, 500 μL of a 2.5 % Glutaraldehyde solution in PBS was added for 1h. Then, gels were washed 3x with 1X PBS and analysed by Confocal microscopy at different depths.

### 2.8. Temporal evaluation of nanocapsules penetration in 3D collagen gels during 12 h

According to the protocol described in Section 2.6, gels were prepared with the same procedure. However, in this case 0.5 mL of the obtained mixture was added to a four wells Chamber Slide$^{TM}$ system, then was delicately removed 300 μL of mixture from each well and it was incubated at 37 °C at an atmosphere of 5 % $CO_2$ overnight. The resulting collagen gels were employed one day after gels formation. Next day, a solution of nanocapsules with a protein concentration (calculated according to protocol described in 2.4 section) of 0.045 mg/mL (corresponds to a theoretical activity of 1.3 U/mL) was incubated in different conditions to obtain three samples: 0.05 M $Na_3PO_4$ pH 7.2 buffer at 70 °C for **nCol-A,** in 0.05 M $Na_3PO_4$ pH 7.2 buffer for **nCol-B**, and in 0.05 M $Na_3PO_4$ pH 5.5 buffer for **nCol-C.** nCol-A was incubated in a lab oven at 70 °C for 1.5 h to get the enzyme inactivated by thermal shock, while nCol-B and nCol-C were kept the same time at 4 °C to preserve their enzymatic activity. Then, a 50 μL of each sample was added on top of these gels, and red and green fluorescence were monitored at different depths in 2 h intervals during 24 h at 37 °C by confocal microscopy.

### 2.9. Statistical analysis

The results shown throughout the article are displayed as mean ± standard error of the mean (SEM), unless otherwise stated. Statistical evaluation of quantitative data was carried out using the Student's T-test. p values < 0.05 were considered to be significant.

## 3. Results and discussion

### 3.1. Labelling of collagenase enzyme with FITC (Col-FITC)

For the correct identification of Collagenase distribution and diffusion throughout a tissue, it is necessary to label the enzyme with an intense fluorophore. It is known by proteomic studies that Collagenase, as most of enzymes, possess several Lysine groups on its tertiary structure. The Fluorescein-labelled Collagenase (Col-FITC) was obtained through the covalent anchoring of the fluorescent dye to the amine groups from Lysine amino acids presents in the enzyme. For this, Fluorescein Isothiocyanate (FITC) dissolved in DMSO was added to a solution of Collagenase in $NaHCO_3$ buffer pH 8.5. It was stirred during 2 hours at RT to originate the carbamate covalent attachment by nucleophilic attack of free amine groups from lysines present in the enzyme to the isothiocyanate group of the fluorophore. The carbamate displays high chemical and proteolytic stability in physiological conditions due to the low electrophilicity of carbonyl group [24]. Finally, the obtained FITC-labelled Collagenase was incorporated into a 10 kDa cut-of filter (Amicon) and was centrifugated and washed several times with buffer to remove the unreacted FITC. By this way, the Col-FITC remains retained in the filter after centrifugation cycles at 3000 rpm for 15 min, while the washing solutions sweep along the unreacted FITC. The green fluorescence was measured after each washing step requiring five washing steps to remove completely the excess of FITC. Then, the labelled protein was collected at 1500 rpm for 5 min. The variation in the surface charge of protein was tested by measuring the Z potential of Col-FITC, which gave a value of – 30.2 mV. Comparing this value with native collagenase (- 7.1 mV) indicates that the incorporation of FITC reduce the amine groups in the surface of the protein and therefore the value become more negative. The correct anchoring of FITC to collagenase was also confirmed in further steps measuring the fluorescence intensity of the nanocapsules.

### 3.2. Synthesis of fluorescein-collagenase nanocapsules (nCol-F)

The incorporation of fluorophores on the nanocapsule surface allows us to monitor their behavior inside the host tissue. The formation of a polymeric nanocapsule around the collagenase preserve it from degradation and therefore, it maintains its enzymatic activity. For the encapsulation of Col-FITC is mandatory to perform the polymerization of monomers in oxygen-free conditions, to prevent the quenching of formed free radicals with the $O_2$ dissolved in the buffer. For this reason, the solution containing Col-FITC was deoxygenated by bubbling $N_2$. The buffer was also deoxygenated with three freeze-thaw cycles under $N_2$ atmosphere to ensure the removal of oxygen in the aqueous buffer.

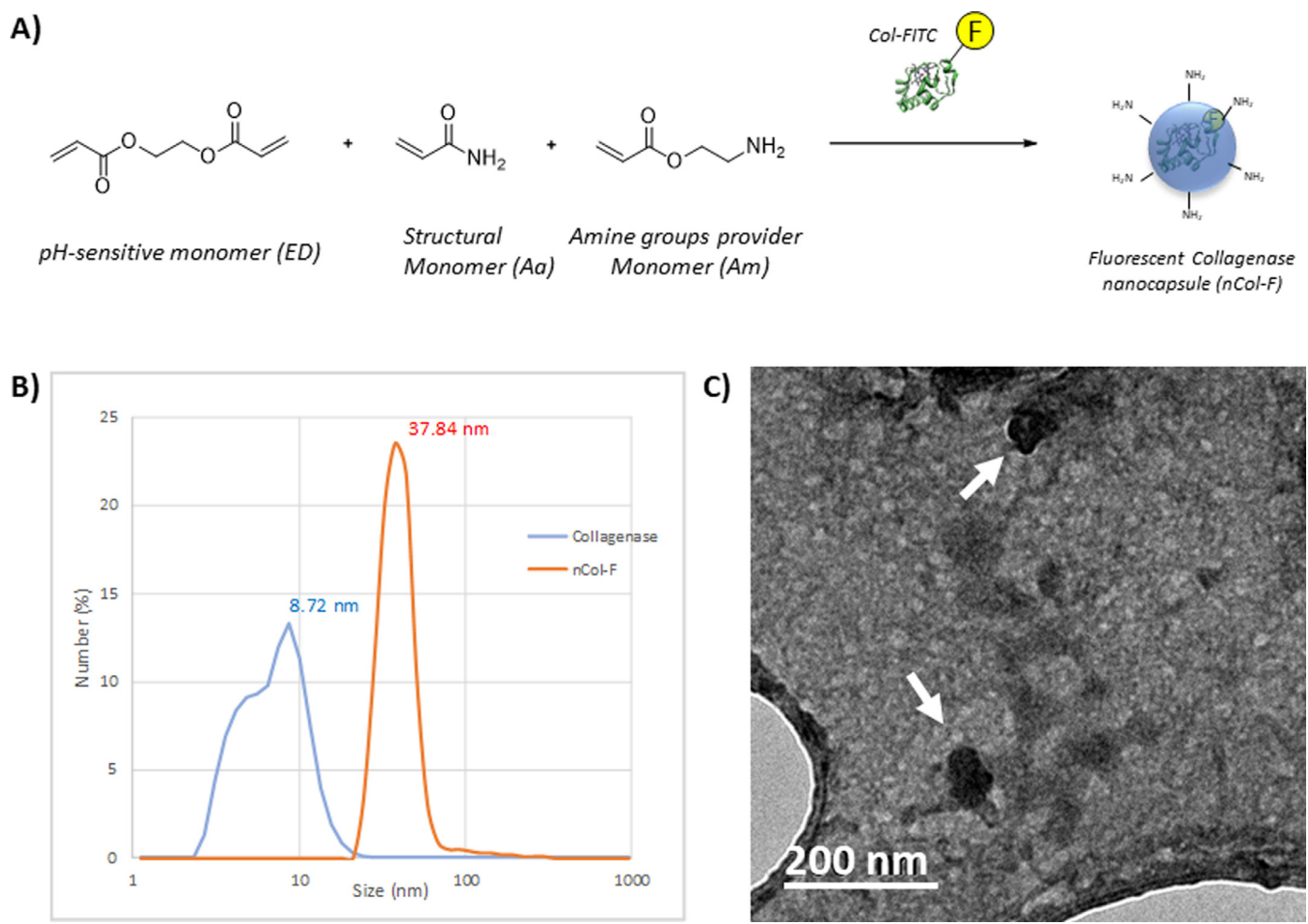

**Fig. 1.** A) Schematic synthetic procedure of collagenase nanocapsules. B) DLS analysis of free collagenase and nCol-F nanocapsules. C) TEM micrography of nCol-F, showing an average diameter of 30-40 nm.

The monomers employed for the nanocapsule formation were acrylamide (Aa) as a structural monomer to complete the whole polymerization and 2-aminoethylmetacrylate hydrochloride (Am) as an amine groups provider monomer. Amino groups provide reactive functional groups for the anchoring of the corresponding fluorophores at the same time that to provide colloidal stability to the nanocapsules in the ionic aqueous medium. Finally, ethyleneglycol dimetacrylate (EG) was employed as pH-sensitive cross-linker, allowing to the capsule to be hydrolysed when the pH is acid, as it has been reported previously [18]. These monomers were dissolved in previously deoxygenated $NaHCO_3$ buffer and were subsequently added to the deoxygenated solution of Col-FITC. This mixture was stirred under $N_2$ atmosphere to get the protein surrounded by monomers. Then, the slowly addition of radical initiators APS and TMEDA initiates the polymerization between monomers which are surrounding the protein. This reaction mixture was stirred at RT under inert atmosphere to complete the polymerization process around the protein yielding the nanocapsules. After this, encapsulated enzymes were purified by centrifugation employing 10 kDa cut-of filters (Amicon) and then, they were washed three times with $NaHCO_3$ buffer to remove the excess of monomers yielding collagenase nanocapsules (nCol-F). A schematic procedure of this synthesis is shown in Fig. 1-**A**. The hydrodynamic diameter of the nanocapsules was measured by Dynamic Light Scattering (DLS), and their size was compared with collagenase before encapsulation. Collagenase shown an average diameter of 8.7 nm, while nCol-F shown a distribution centred in 37.8 nm (Fig. 1-**B**). This size is consistent with the average size obtained by TEM analysis of around 30-40 nm (Fig. 1-**C**). Zeta potential measurement changed from – 30.2 mV of Col-FITC to a value of – 14.1 mV, which became less negative due to the presence of amine groups in the nanocapsules surface.

### 3.3. Rhodamine-labelling of nCol-F (nCol-F-Rh) and nanocapsules activity measurements

For the study of nanocapsules hydrolysis process, collagenase was labelled with FITC for the correct protein identification. These Col-FITC were encapsulated, and the obtained nanocapsules (nCol-F) were labelled with another fluorophore for their visualization by fluorescence microscopy throughout the gel. The labelling mechanism is the same as mentioned before; the correspondent fluorophore-isothiocyanate react with amine groups from the polymeric framework of the nanocapsules, yielding a stable carbamate conjugate. For our purpose, the nanocapsules were labelled with

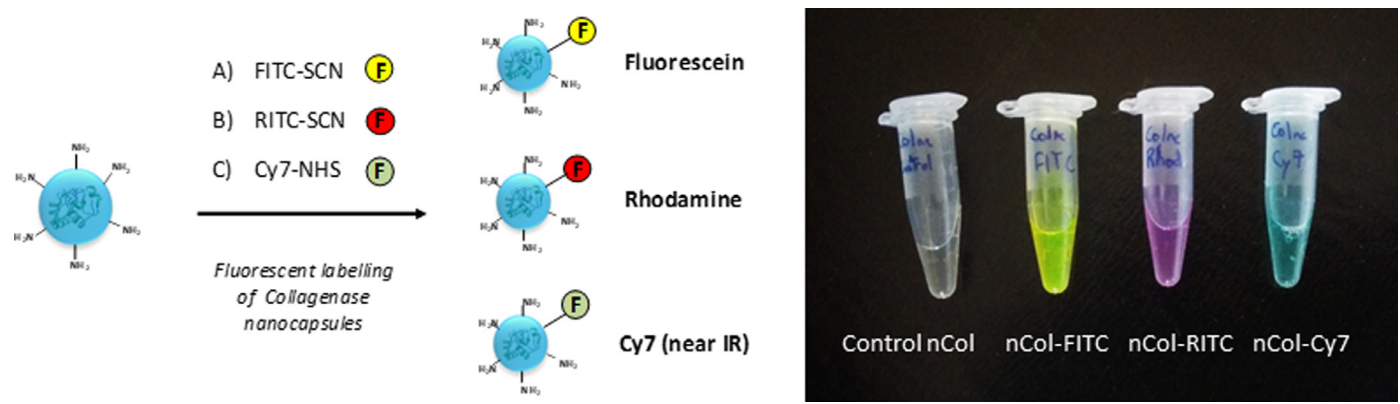

**Fig. 2.** Left: Schematic procedure for the labelling of Collagenase nanocapsules with FITC, RITC and Cy7. Right: Image of a control Collagenase nanocapsules and Fluorophores-labelled Collagenase nanocapsules.

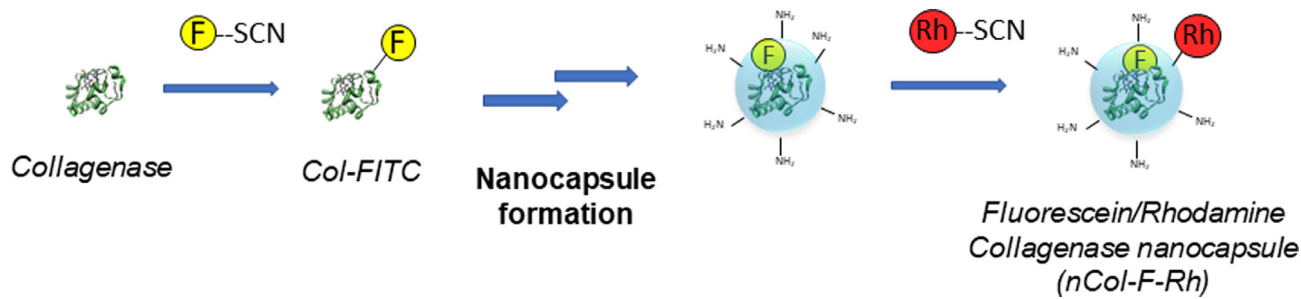

**Fig. 3.** Schematic representation of double-fluorescent labelling steps of Collagenase nanocapsules.

Rhodamine Isothiocyanate (RITC), which provides a red fluorescence to the nanocapsules without interference with the green fluorescence delivered by Col-FITC. However, it is noteworthy that other fluorophores can be used such as FITC or Cyanine7-NHS ester (Cy7). In three cases, the fluorescence was kept after several washing steps, as is showed in Fig. 2. These fluorophores could be used indistinctly, depending on the desired wavelength.

Dual red/green fluorescence in Collagenase nanocapsules (nCol-F-Rh) were synthesized employing RITC as fluorophore following a similar procedure to the employed in FITC grafting. After that, the obtained nCol-F-Rh were purified by centrifugation employing 10 kDa Amicon following the same steps carried out in the first labelling to remove the excess of RITC. A short scheme of the double-labelling process is shown in Fig. 3. The Z potential of nCol-F-Rh was measured, and the value increased from − 14.1 mV of nCol-F to − 7.8 mV. This change in the surface charge is consistent with the incorporation of RITC group on the surface of nanocapsules, which is positively charged at pH of 8.5.

The encapsulation of collagenase should reduce its enzymatic activity because the encapsulated enzyme cannot access to the substrate (collagen) due to the presence of the polymeric shell. It is necessary to release the enzyme from the polymeric nanocapsule to restore its capacity to digest the collagen matrix. The transient enzymatic activity loss of the encapsulated collagenase was determined measuring the proteolytic capacity of nanocapsules in comparison with free collagenase which suffered the same purification steps but without being encapsulated within a polymeric shell. The enzymatic activity of the collagenase trapped in the collagenase nanocapsules can be reduced due to the friction forces and temperature increases required in the numerous purification steps required in the synthesis and this reduction should be discarded to the reduction produced by the encapsulation process (Fig. 4A). In-terestingly, the enzymatic activity of free collagenase was rapidly reduced in mild acidic conditions whereas the encapsulated collagenase maintained its proteolytic capacity during longer times. Thus, the encapsulated enzyme retained more than 40 % of enzymatic activity after 24 hours at pH = 5.5 in comparison with free enzyme which showed negligible activity (Fig. 4B). In this graphic, the enzymatic activity of free collagenase was normalized to the corresponding 30-40 % enzyme activity in nanocapsule form with a correction factor of 0.4 to avoid the overestimation of the enzymatic activity of free collagenase.

It is noticeable that the decrease of enzymatic activity during nanocapsules formation and labelling of fluorophores processes is around a 70 % comparing with the activity of native collagenase. This drastic reduction was not permanent while it was recovered once the nanocapsules were hydrolysed as it is showed in Fig. 4B. The polymeric nanocapsule acts as a shield that protect the enzyme but also cover the active site. It justifies the high decrease in the enzymatic activity when the enzyme is encapsulated. Nevertheless, as it was reported in previous works, once the nanocapsule is degraded by the effect of the pH, the enzymatic activity is mainly recovered [18]. Both samples were stored at 4° C during 24 h and their activity was newly measured, but there are no significant change in their enzymatic activities. Free enzyme and nCol-F-Rh preserves their activity at low temperatures.

It is also important to ensure that the fluorescence capacity of fluorescein and rhodamine is not affected significatively during the synthesis process. For this reason, it was measured the red and green fluorescence emitted from *nCol-F-Rh,* and data obtained were standardized to 1 mg of nanocapsules. For green fluorescence was obtained a concentration value of 16,9 μg of FITC in 1 mg of protein. In the case of Red Fluorescence of nCol-F-Rh, it was obtained a value of 172,8 μg of RITC in 1 mg of protein. This value is around 10-fold higher than the fluorescence derivate from FITC, due to the high availability of amine groups in the surface of the nanocapsules comparing with the amine groups present in the enzyme.

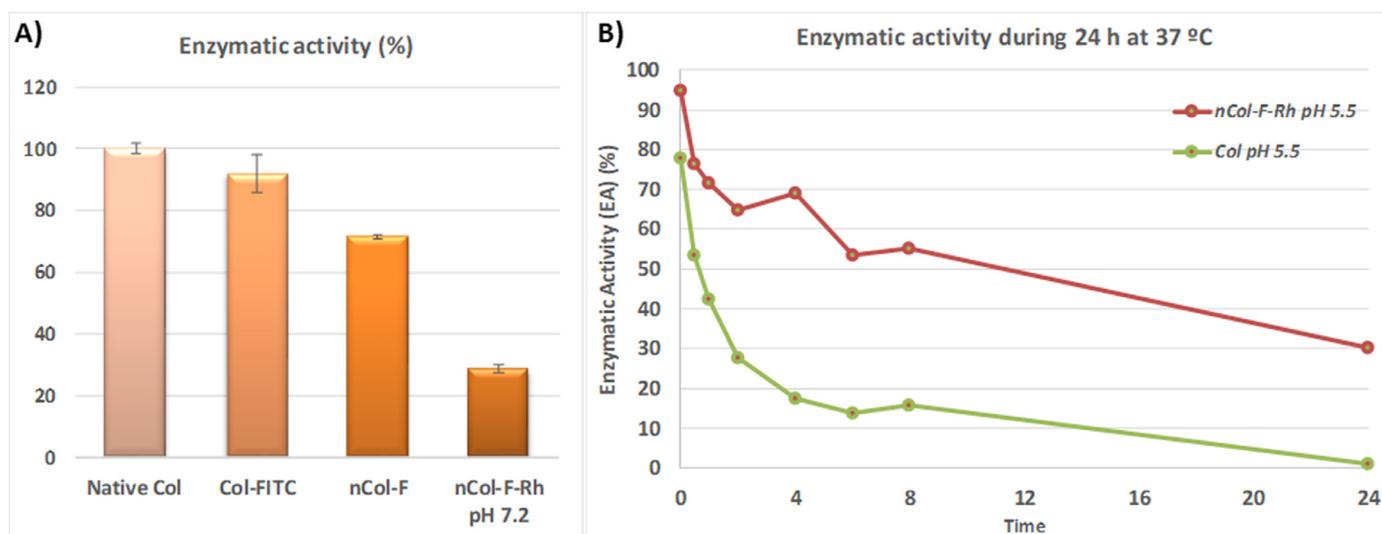

**Fig. 4.** A) Enzymatic activity after each synthetic step, B) Enzymatic activity of collagenase nanocapsules and collagenase at mild acidic conditions.

### 3.4. Evaluation of the penetration of nCol-F-Rh nanocapsules in 3D collagen gels

In order to estimate the capacity of collagenase to penetrate into collagen gels designed to mimic an extracellular matrix, nCol-F-Rh were incubated in a three-dimensional Collagen gel at different pH to promote the hydrolysis of the pH sensitive red-labelled polymeric shell of nanocapsules and the subsequent diffusion of Col-FITC enzyme retained inside due to the continuous digestion of the collagen from the gel. Both end-point penetration experiment and temporal evaluation penetration were performed for the evaluation of colour differences inside the gels over time.

#### 3.4.1. Nanocapsules penetration evaluation in a 18 h end-point experiment

One day after the formation of gels, the resulting 3D collagen gels were employed for the nCol-F-Rh penetration experiment. For this purpose, two samples of nanocapsules solution (2,57 U/mL) were prepared at pH 7.2 and pH 5.5. Then, they were added on the centre of the collagen gel and they were incubated for 18 h. Next day, gels were fixed with a 2.5 % Glutaraldehyde solution in PBS to stop the collagen degradation and to protect the fluorescence capacity of fluorophores. Then, gels were washed with PBS and analysed by Confocal microscopy at different depths, obtaining the XY planes superposition in the Z-axis for different samples at physiological pH, mild-acidic conditions and gel without nCol-F-Rh as control. The obtained representations can be observed in Fig. 5. Due to the pH-sensitivity nature of the collagenase nanocapsules, at pH 5.5 the hydrolyzation of the pH-sensitive monomers present in the polymeric shell was faster than the observed at physiological pH. Mild-acidic conditions provoked rapid release of the Col-FITC from nanocapsules and allowed to the enzyme to digest the collagen matrix of the gel. As result of this, higher penetration rate was observed at acid pH in comparison with physiological pH. This process is schematized in Fig. 5-A. Collagenase labelled with FITC were more homogeneously distributed in the entire gel reaching deeper zones in mild acidic conditions than in the case of nanocapsules incubated at physiological pH (Fig. 5C and 5D). It is observable that red fluorescence (represented in purple) derived from the labelling of the polymeric shell of nanocapsules was slightly more abundant inside the gel in the sample at pH 5.5 in comparison with pH 7.2. The triggered release of Col-FITC at acid pH allowed a fast digestion of collagen from gel and therefore, nanocapsules were capable to penetrate deeper in the gel than the nanocapsules exposed to physiological pH. The control gel did not show autofluorescence derived from the collagen (Fig. 5B).

#### 3.4.2. Temporal evaluation of nCol-F-Rh nanocapsules penetration in 3D collagen gels for 12 hours

For a precise temporal analysis of the nanocapsules hydrolysis progress at physiological and acidic pH, the experiment was performed in a four wells Chamber Slide™ system, which was then coupled to a confocal microscope for gel analysis. Gels were prepared, as described in Section 2.8, inside this Chamber slide. Then, it was inserted in a humid chamber to prevent the dehydration of the gels (thereby avoiding the loss of its volume) and the temperature was kept at 37 °C for an optimal enzyme activity and also to assure the same environmental conditions between samples. This assembly allowed us to analyse sequentially the samples every 2 hours for a depth of 50 μm in each gel during a period of 12 hours. For this, three samples were incubated during 1.5 h in different conditions. One of them correspond to inactive nCol-F-Rh (**nCol-A**) which was incubated at 70°C and pH 7.2 to eliminate the enzymatic activity by thermal denaturalization. Other sample was nCol-F-Rh incubated at pH 7.2 (**nCol-B**) to emulate physiological conditions. Finally, the last sample was nCol-F-Rh incubated at pH 5.5 (**nCol-C**). 50 μL of each sample were added on top of gels, and then, red and green fluorescence were monitored for each sample. Gels were analysed and the projected fluorescence signals in the ZY plane were represented to appreciate the nanocapsule and collagenase penetration in each case. The penetration depth of each gel each 2 h of analysis can be observed in the videos from ***Supporting Information (S1 for nCol-A; S2 for nCol-B and S3 for nCol-C).***

The comparative analysis between nCol-F-Rh at pH 7.2 and pH 5.5 in Fig. 6-**A** show that red fluorescence was present in the upper layers of the gel at both physiological and acidic pH after 12 h of experiment. However, in the case of green fluorescence in Fig. 6-**B,** it is observable that after 12 h of experiment, the fluorescence front reached deeper regions of the gel at pH 5.5 comparing with pH 7.2. This fact suggests that nanocapsules exposed to mild acidic conditions were hydrolysed in higher amount than the ones exposed to physiological pH and therefore, they release higher amount of collagenase to the media which were able to diffuse to inner zones of the gel. Interestingly, polymeric nanocapsules could not penetrate deeply into the tissue even in the case of the ones exposed to mild acidic conditions. Polymeric nanocapsules maintain their integrity due to the presence of the pH-sensitive cross-linkers which could be not completely hydrolysed in this period of time, even in the case of mild-acidic conditions. Therefore, the hydrolysis of the polymeric framework in this last condition is enough to release the collagenase but not enough to allow a free diffusion of the hydrolysed polymer chains into the gel.

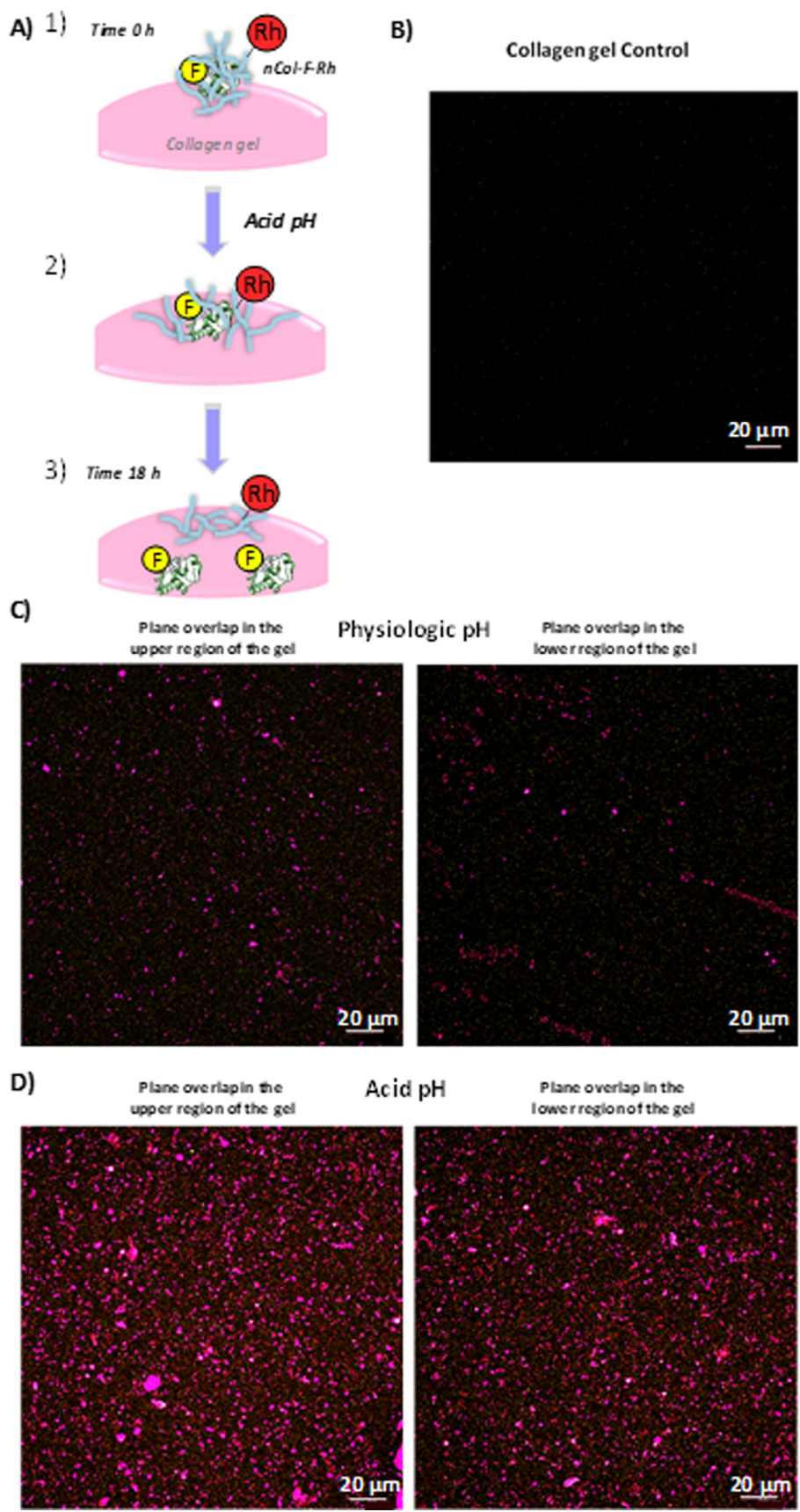

**Fig. 5. A)** Schematic representation of nCol-F-Rh degradation inside the collagen gel. **B)** Control Collagen gel without nCol-F-Rh. **C) and D)** Confocal images Z-axis superposition of Collagen gel incubated with nCol-F-Rh for 18 h at physiological pH (C) and at Acid pH (D) for upper and lower fronts of the gel. Observation depth in each case were 300 $\mu$m for C and 320 $\mu$m for D, respectively. Images were taken each 2 $\mu$m. Scale bar: 20 μm.

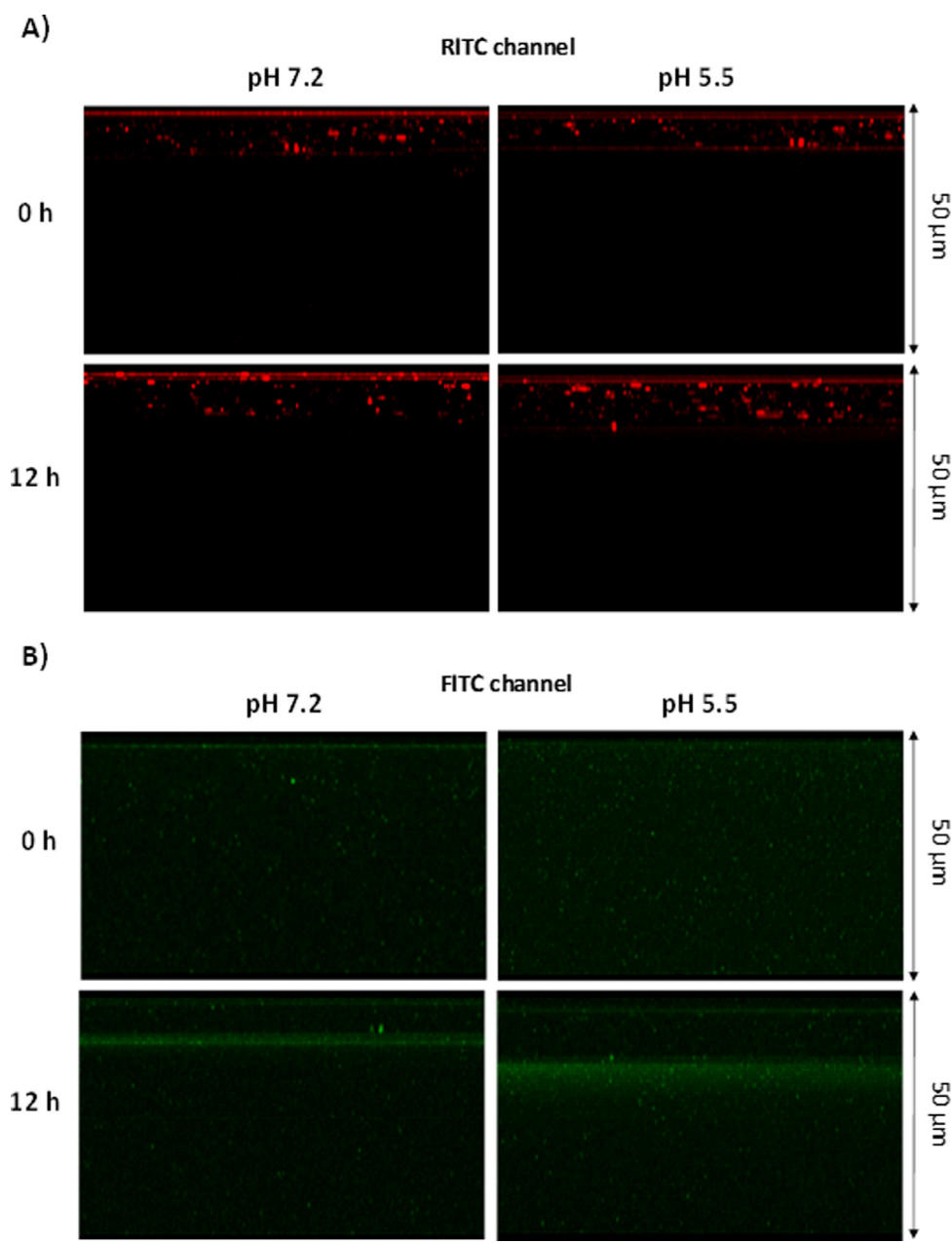

**Fig. 6.** Comparative confocal images of a Z-axis projection of collagen gel incubated with nCol-F-Rh at physiological pH (Left images) and acidic pH (Right images) at times 0 and 12 h. **A)** Images correspond with Red fluorescence related to RITC. **B)** Images correspond with Green fluorescence of FITC. All gels were compared for a depth of 50 μm.

At physiological pH of 7.2, the temporal progression of green fluorescence across the inner gel is slower comparing with the progression in the case of enzyme previously incubated at pH of 5.5. In this case, the distribution of the fluorescent "cloud" is much wider along the gel and comparatively higher than the case of pH 7.2, indicating that the premature hydrolysis of the nanocapsules allowed the collagenase to diffuse freely inside the Collagen gel and thereby, digested the collagen achieving higher penetration rates in comparison with the nanocapsules incubated at physiological pH.

To obtain a detailed profile for the evolution of red and green fluorescence along the gel, the raw fluorescence data of nCol-F-Rh in different conditions (**nCol-A, B and C**) was extracted and analysed. The green fluorescence data were represented in an Area graph, overlaying all the Z planes of the gel to obtain a Z projection of the whole gel. In that way, it is represented the fluorescence intensity area *vs* gel depth for each time. Each time of analysis are overlaid in the graph to appreciate the progress of depth *vs* time in the sample. Analysing the green fluorescence (relative to Col-FITC) for the case of **nCol-A** (Fig. 7-**A),** it was observable that the fluorescence intensity was concentrated in first 4 μm of the gel, and the intensity profile did not change over time. This ineffective penetration indicated that the nanocapsules have lost their enzy-matic activity with the thermal shock and there was no digestion of the collagen. In the case of **nCol-B** sample (Fig. 7-**B**), it was observable that the fluorescence intensity profile has slightly progress comparing with inactive enzyme, reaching a fluorescence front between 9-15 μm depth at 8 h of analysis, with a poor increase of the depth at the time of 12 h. This effect can be attributed to a very slow degradation of nanocapsules at pH 7.2 which provoked a poor collagen degradation, and consequently, poor penetration of fluorescent collagenase. Finally, in **nCol-C** sample (Fig. 6-**C**), it is noteworthy that the fluorescence intensity profile has considerably advanced to deeper zones in the gel in comparison with nanocapsules at pH 7.2. In this case, at 4 h of experiment the whole fluorescence front was situated between 13-20 μm depth of the gel. Then, the front advances over time, achieving a wide distribution of the fluorescence between 17 and 30 μm depth after 12 h of analysis. The faster degradation of nanocapsules at pH 5.5 allowed the fluorescent enzyme to rapidly degrade the collagen, obtaining two-fold enzyme penetration ratios for the same time, and with very low enzyme concentrations.

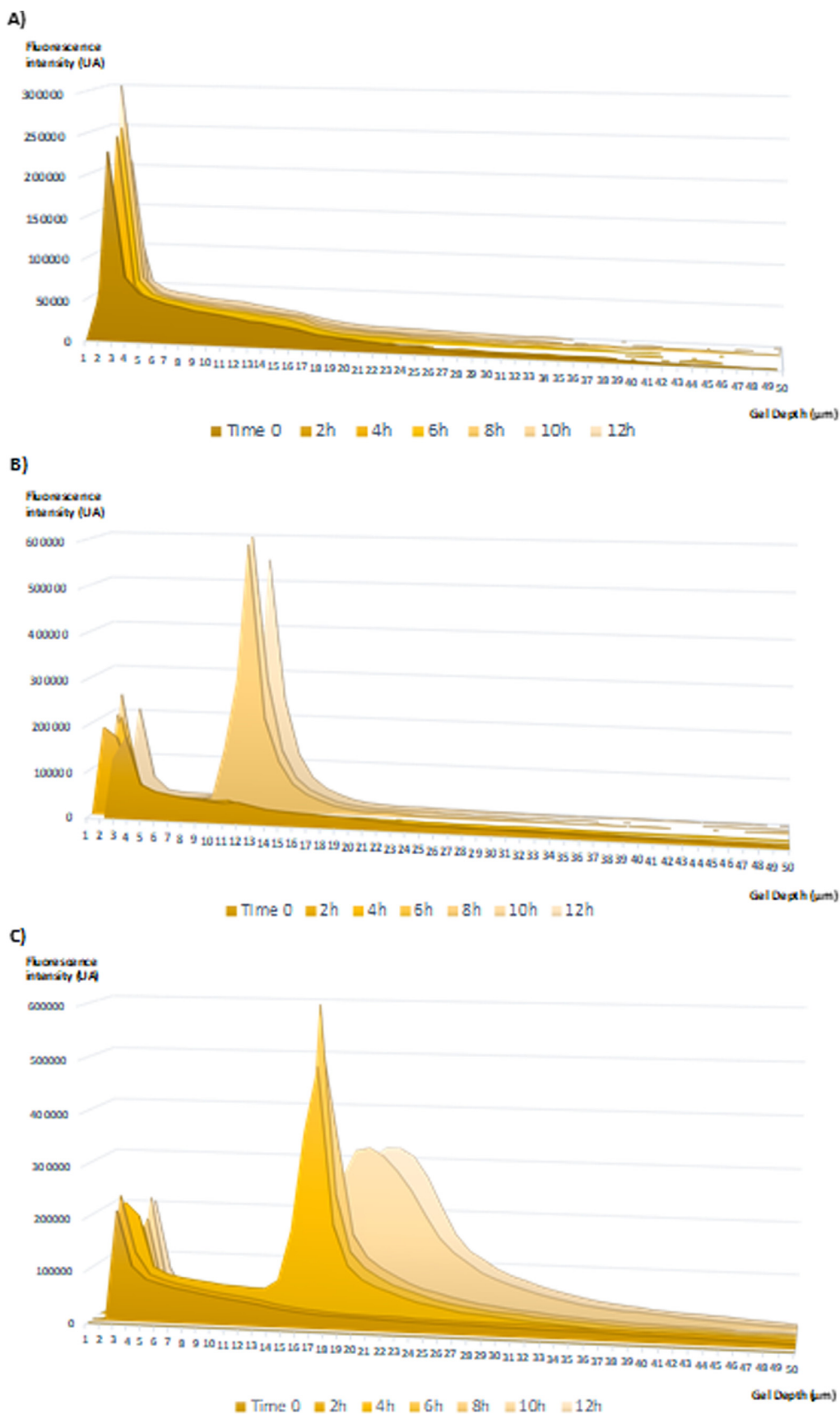

**Fig. 7.** Green Fluorescence intensity histograms for each gel layer analysed every 2 hours. **A)** Histogram correspondent to inactivated nCol-F-Rh (**nCol-A**). **B)** Histogram correspondent to nCol-F-Rh preincubated at pH 7.2 (**nCol-B**). **C)** Histogram correspondent to nCol-F-Rh preincubated at pH 5.5 (**nCol-C**).

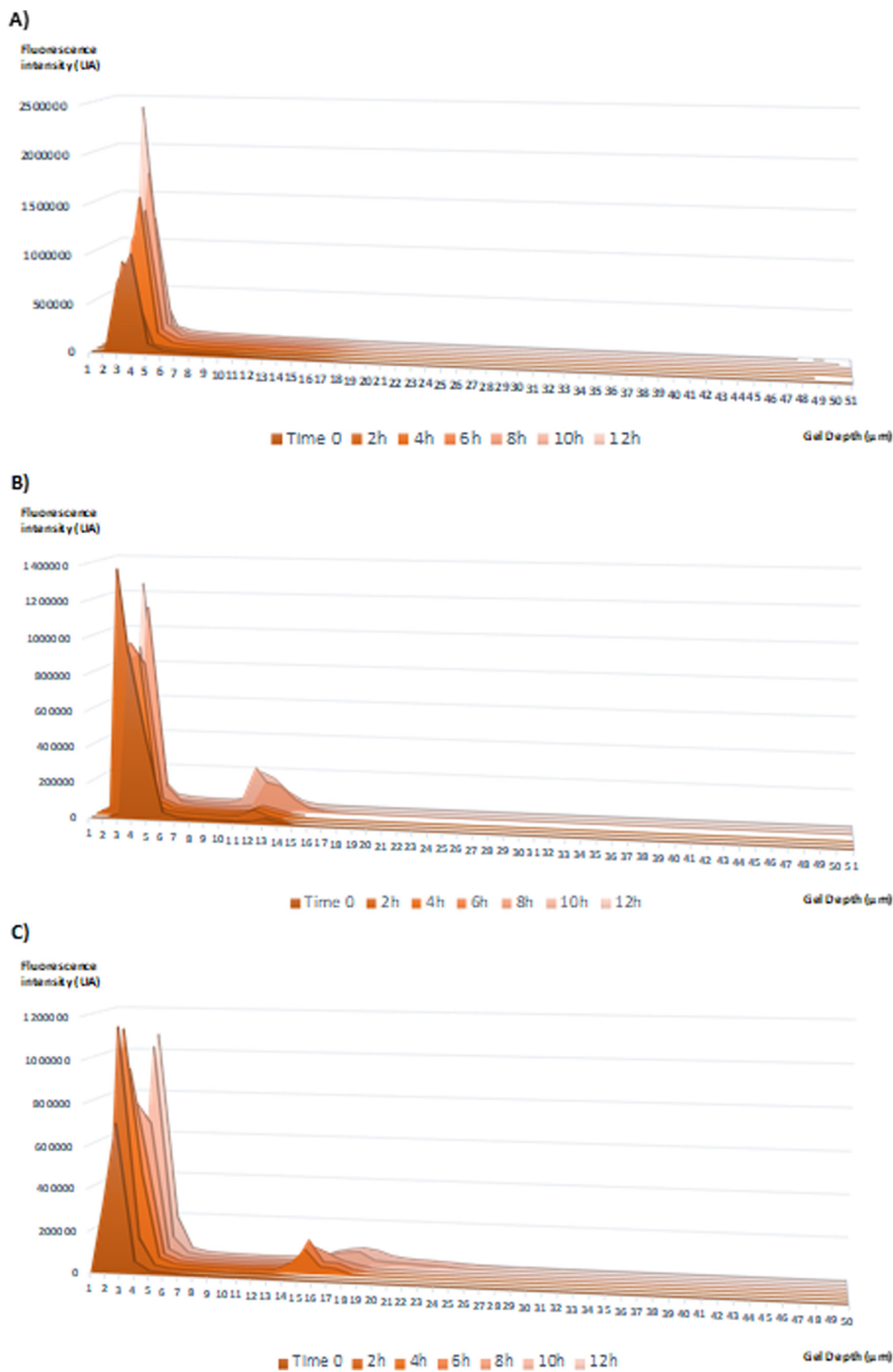

**Fig. 8.** Red Fluorescence intensity histograms for each gel layer analysed every 2 hours. **A)** Histogram correspondent to inactivated nCol-F-Rh (**nCol-A**). **B)** Histogram correspondent to nCol-F-Rh preincubated at pH 7.2 (**nCol-B**). **C)** Histogram correspondent to nCol-F-Rh preincubated at pH 5.5 (**nCol-C**).

Through the analysis of the histograms for red fluorescence intensity (relative to nanocapsules polymeric structure) it was observed that the fluorescence intensity was limited to the first 5 μm of the collagen gel with **nCol-A** (Fig. 8A) whereas in the case of **nCol-B** (Fig. 8B) and **nCol-C** (Fig. 8C) only a little fraction of polymeric nanocapsules were able to penetrate deeply into the gel: 11-13 μm at physiological pH and 14-19 μm, in the case of nanocapsules incubated in mild-acidic conditions. Most of the red fluorescence dots were concentrated in the first 5 μm of the gel in both cases.

These results suggest that nanocapsules did not penetrate by themselves remaining in the close surface of the gel. The reason of this poor penetration can be that, even in the case of nanocapsules exposed to mild acidic conditions, the nanocapsules retained a certain amount of non-hydrolysed crosslinkers which partially maintained their integrity and therefore, their penetration into the gel were compromised. As result of this fact, only a low amount of red fluorescent dots (which correspond to unaltered or hydrolysed nanocapsules) were detected in deep zones of the gel. In any case, the presence of these red dots in deep zones of the gel are higher in the case of nanocapsules exposed to mild acidic conditions due to the higher release of collagenase in the zone. This also reinforce the idea that nanocapsules present low activity before degradation due to the presence of polymer shell that act as a "shield" which protect the enzyme, as has been mentioned above. However, the premature degradation of nanocapsules at pH 5.5 release the enzyme from the polymeric framework, recovering the enzymatic activity and reaching higher penetration ratios of green fluorescence dots due to the presence of FITC-labelled enzyme. Thus, the nanocapsules acted as collagenase reservoir that, at mild acidic conditions, released collagenase which can diffuse to deep zones of the tissue.

## 4. Conclusions

In conclusion, a new strategy to evaluate the hydrolysis and penetration of pH-sensitive collagenase nanocapsules is presented. For this, the collagenase nanocapsules were labelled with two different fluorophores, fluorescein and rhodamine. Fluorescein was attached to native collagenase enzyme while rhodamine was anchored on the nanocapsule surface. The enzyme activity was altered in the fluorophore anchoring and polymer coating processes but did not permanently compromise its functionality. The double labelling allowed us to evaluate the penetration capacity of collagenase nanocapsules, and additionally the diffusion capacity of collagenase enzyme once the polymeric framework of nanocapsules was hydrolysed. Due to automatic-confocal microscopy analysis it was assured a complete reproducibility of procedure conditions between samples. As a result, it was observed that Col-FITC diffusion capacity in acidic pH was higher than in the case of physiological pH, due to the pH-responsiveness nature of these nanocapsules. It was also analysed in detail the evolution of fluorescence over time inside the gel, obtaining that the enzyme was capable to penetrate two-fold deeper in acidic medium than in physiological medium in 12 hours, even though the enzyme concentration was low. Meanwhile, the red fluorescence of polymeric shell was retained in both cases in the first few layers and did not penetrated into the tissue. These collagenase nanocapsules displayed a high enzymatic activity at low concentrations and their optimal operation opens a wide range of possible nanomedical applications.

## Supporting Information

Video of the penetration of nCol-F-Rh at pH 7.2 and 5.5 in Collagen gels.

## Declaration of Competing Interest

The authors declare that they have no known competing financial interests or personal relationships that could have appeared to influence the work reported in this paper.

## Acknowledgments

Author Contributions: The manuscript was written through contributions of all authors. All authors have given approval to the final version of the manuscript. The authors declare no competing financial interest. This work was supported by European Research Council; ERC-2015-AdG (VERDI), Proposal No. 694160 and by ERC-2018-PoC (N° 825435) DECOMPACT.

The authors wish to thank the ICTS Centro Nacional de Microscopia Electrónica (Spain) and CAI Cytometer and Fluorescence microscopy of the Universidad Complutense de Madrid (Spain) for the assistance and Servier Medical art for the Creative Commons figures.

## Supplementary materials

Supplementary material associated with this article can be found, in the online version, at doi:10.1016/j.actbio.2020.12.022.